\documentclass{aa}
\usepackage{graphicx}
\usepackage{times}
\begin{document}

\title{High-precision elements
of double-lined spectroscopic binaries from combined
interferometry and spectroscopy}
\subtitle{Application to the $\beta\,$Cephei star 
$\beta$\,Centauri\thanks{Radial-velocity data available electronically
from the CDS via anonymous ftp to cdsarc.u-strasbg.fr (130.79.128.5)}}

\author{M.\ Ausseloos\inst{1},
C.\ Aerts\inst{1,2},
K.\ Lefever\inst{1},
J.\ Davis\inst{3}, \and
P.\ Harmanec\inst{4,5}}

\institute{Instituut voor Sterrenkunde, Katholieke Universiteit Leuven,
Celestijnenlaan 200 B, B-3001 Leuven, Belgium \and
Department of Astrophysics, Radboud University Nijmegen, PO Box 9010, 6500 GL
Nijmegen, the Netherlands\and
Chatterton Astronomy Department, School of Physics, University of Sydney,
NSW 2006, Australia\and
Astronomical Institute of the Charles University, V
Hole\v{s}ovi\v{c}k\'ach~2, CZ-180~00~Praha 8, Czech Republic \and
Astronomical Institute, Academy of Sciences, CZ-251~65~Ond\v{r}ejov, Czech
Republic}

\authorrunning{Ausseloos et al.} \titlerunning{Fundamental parameters and
  oscillation frequencies of the SB2
  $\beta\,$Cen}

\abstract{}{We present methodology to derive high-precision
estimates of the fundamental parameters of double-lined
spectroscopic binaries.
{We apply the methods to the case study of the double-lined $\beta\,$Cephei
star $\beta\,$Centauri. We also present a detailed analysis of 
$\beta\,$Centauri's
line-profile variations caused by its oscillations.}} 
{High-resolution spectral time series and visual or interferometric
data with a good phase distribution along the orbital period are required.
We point out that a
systematic error in the orbital amplitudes, and any quantities
derived from them, occurs if the radial velocities of blended
component lines are computed without spectral disentangling. This
technique is an essential ingredient in the derivation of
the physical parameters if the goal is to obtain a precision of
only a few percent. We have devised iteration schemes to obtain
the orbital elements for systems whose lines are blended
throughout the orbital cycle.}  {We derive the
component masses and dynamical parallax of $\beta\,$Centauri
with a precision of 6\%
and 4\%, respectively. Modelling allowed us to refine the mass
estimates to 1\% precision resulting in $M_1=10.7\pm 0.1\,M_\odot$
and $M_2=10.3\pm 0.1\,M_\odot$, and to derive the age of the system
as being $(14.1\pm 0.6)\times 10^6$\,years.
{We deduce two oscillation frequencies for the broad-lined primary of 
$\beta\,$Centauri: 
  $f_1=7.415\,$c\,d$^{-1}$ and $f_2=4.542$\,c\,d$^{-1}$ or one of their 
  aliases. The degrees of these oscillation modes are higher than 2 for both
  frequencies, irrespective of the alias problem. No evidence of oscillations in
  the narrow-lined secondary was found.  }}  
{We propose that our
iteration schemes be used in any future derivations of the
spectroscopic orbital parameters of double-lined binaries with
blended component lines to which disentangling can be successfully
applied. The combination of parameters resulting from the
iteration schemes with high-precision estimates of the orbital
inclination and the angular semi-major axis from interferometric or
visual measurements allows a complete solution of the system.}

 \maketitle

\section{Introduction}
\label{cen_intro}

Binary stars have long been considered as astrophysical
laboratories, providing one of the best tests of stellar structure
models (e.g., Maceroni 2005 and Ribas 2005, for recent reviews).
This is particularly so for massive binaries because their
structure and evolution are not well-understood, while being of
great importance for the chemical enrichment and evolution of
galaxies. The convective and rotational mixing properties of
massive stars with a well-developed convective core are still
poorly calibrated, while being the dominant factors determining
their evolution (e.g.\ Maeder \& Meynet 2000).  While
observational capabilities and analysis tools have improved
significantly in recent years (e.g.\ Hilditch 2004a), there is
still a lack of ultra-precise fundamental parameter determinations
of binaries with an OB-type component (e.g.\ Hilditch 2004b).
Indeed, component mass estimates with a precision better than 2\%
are available for relatively few such systems, although such a
precision is necessary to provide stringent observational tests
for stellar structure and evolution models (e.g.\ Andersen 1991).
In this paper, we provide methodology to achieve a high precision
for mass estimates from combined interferometric and spectroscopic
data of double-lined spectroscopic binaries with merged component
lines, and we apply it to the massive binary $\beta\,$Cen. The
methods are based on spectral disentangling {(Hadrava 1995, 1997, 2001,
  2004b).} 

{The bright star $\beta\,$Cen (HD\,122451, HR\,5267, B1\,III, $m_V=0.6$) has
been known to be variable in velocity since the beginning of the twentieth
century. It is the brightest member of the class of $\beta\,$Cephei stars, a
homogeneous group of oscillating B0--B3 stars (see Stankov \& Handler 2005 for a
recent review).  They have low-degree, low-order pressure and gravity modes with
periods of a few hours excited by the $\kappa\,$mechanism (Pamyatnykh
1999). They reveal amplitudes of several tens of mmag down to the detection
threshold in UBV, so these stars are good potential targets for in-depth seismic
studies.  Asteroseismology of $\beta\,$Cephei stars indeed received a lot of
attention lately, after it became clear that their oscillations cannot be
explained in terms of standard evolution models. For two prototypical class
members, the oscillations revealed differential internal rotation and the
occurrence of core convective overshooting (HD\,129929: Aerts et al.\ 2003 and
Dupret et al.\ 2004; $\nu\,$Eri: Pamyatnykh et al.\ 2004 and Ausseloos et al.\
2004).  }

As one of the brightest stars in the southern hemisphere as a whole,
$\beta\,$Cen has been the subject of numerous studies. We refer to Ausseloos et
al.\ (2002, hereafter Paper\,I), Davis et al.\ (2005, hereafter Paper\,II), and
references in these two papers for an overview of these studies, without
repeating all of the information here. We summarise only briefly the
characteristics of the system that are relevant for our current work.

High-resolution spectra covering 12 years revealed that
$\beta\,$Cen is a double-lined spectroscopic binary with an
orbital period of 357 days and an eccentricity of about 0.81
(Paper\,I). Interferometric data assembled with the Sydney
University Stellar Interferometer {(SUSI)} and covering 7 years led to
similar values for the period and eccentricity and, moreover, to
an orbital inclination of $67.4^\circ$ and an angular semi-major axis
of $0.0253''$, as well as to a brightness ratio of 0.868$\pm 0.015$
(Paper\,II).  The spectroscopic variability is due not only to the
binarity, but also to oscillations of the components with periods
of several hours (Paper\,I). The star is photometrically
constant at the level of mmag.

The independent orbital fits to the spectroscopic and
interferometric data have four parameters in common: the orbital
period ($P_{\rm orb}$), the epoch of periastron passage ($E_0$),
the eccentricity ($e$), and the longitude of periastron ($\omega$).
The values of these four differed by less than the 1$\sigma$
uncertainties such that a fully consistent orbital solution for
$\beta$\,Cen was achieved (Paper\,II).  The combination of all the
available information subsequently led to the conclusion that the
system consists of components with equal masses of $M_1=M_2=9.3\pm
0.3\,M_\odot$ (i.e., a precision of 3.2\%) and that it has a
dynamical parallax of 9.78$\pm 0.16$\,mas.

We show in this work that the masses of the components of
$\beta\,$Cen were significantly underestimated due to a systematic
error in the amplitudes of the spectroscopic orbit. This is a
consequence of the inappropriate way in which the radial
velocities were estimated from line profiles of merged spectra. A
similar conclusion was reached recently by Tango et al.\ (2006)
for the triple system $\lambda\,$Sco and occurs for {\it any\/}
spectroscopic binary in which both components contribute to the
lines used for the orbital radial-velocity (hereafter abbreviated
as RV) determination.  Earlier attempts to avoid such systematic
error can be found in Tomkin et al.\ (1995) for the
$\delta\,$Scuti star $\theta^2$\,Tauri. In that work, the authors
subtracted the lines of the primary by means of spectra of
reference stars with the same spectral type before computing the
secondary's RV values. We provide analysis schemes based on
spectral disentangling to overcome this problem of systematic
errors in a more accurate way. Our schemes allow us to eliminate
the systematic errors in the physical parameters. We illustrate
our method by its application to the case of $\beta\,$Cen. Our
methodology is applicable to the analysis of any spectroscopic
binary whose line profiles can be successfully disentangled. It
leads to a significant improvement in the precision of the
physical parameters and dynamical parallax of such systems, 
similar to the case of binaries with emission-line  stars 
(Harmanec 2002).

\section{Methodology for orbital determination}
\label{methods}

The data used to illustrate our methodology are the SiIII
$\lambda$\,4552.6\,\AA\ line profiles of $\beta\,$Cen obtained
over 12 years with the ESO CAT telescope and with the Swiss
Euler telescope, as described in Paper\,I.
\begin{figure}
\centering
\rotatebox{-90}{\resizebox{6.0cm}{!}{\includegraphics{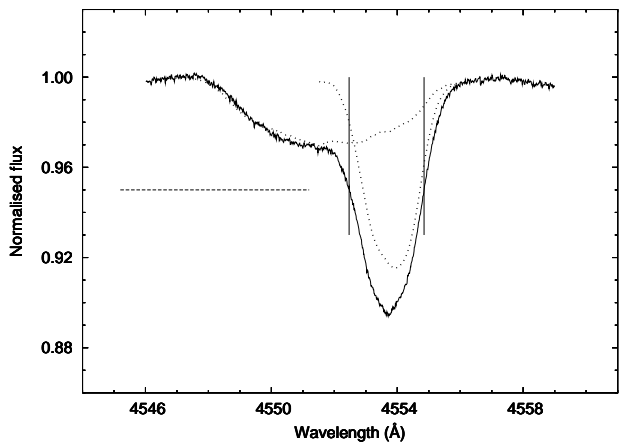}}}
\caption{SiIII $\lambda$ 4552.6\,\AA\ line profile obtained by
averaging ten spectra measured on 16 May 1988 (solid line). The two
vertical lines denote the integration limits which were used in
Paper\,I to calculate the RV of the secondary.  The two dotted
lines denote the disentangled line profiles of both components.
}\label{fig_cen_syst_error}
\end{figure}
By means of illustration of the system's lines and of the occurrence of a
systematic error in the orbital RV determinations used in Papers\,I and II, we
recall here in quite some detail the way the RVs were obtained.  We show in
Fig.\,\ref{fig_cen_syst_error} a typical line profile of the system averaged
over one night of data, i.e., a profile in which the oscillatory variations are
averaged out. As can be seen, the lines produced by the two components are
blended with each other. In none of the available spectra are the two
components' lines well separated. Moreover, the oscillations induce deviations
from a Gaussian shape for each of the line components (see Fig.\,1 in Paper\,I).
For this reason, the RV values of the component with the narrowest line were
derived from the line centroid (first moment, see Aerts et al.\ 1992 for a
definition) in Paper\,I, with the integration limits indicated in
Fig.\,\ref{fig_cen_syst_error} corresponding to a normalised flux value of 0.95.

\begin{figure}
\centering
\rotatebox{-90}{\resizebox{6.0cm}{!}{\includegraphics{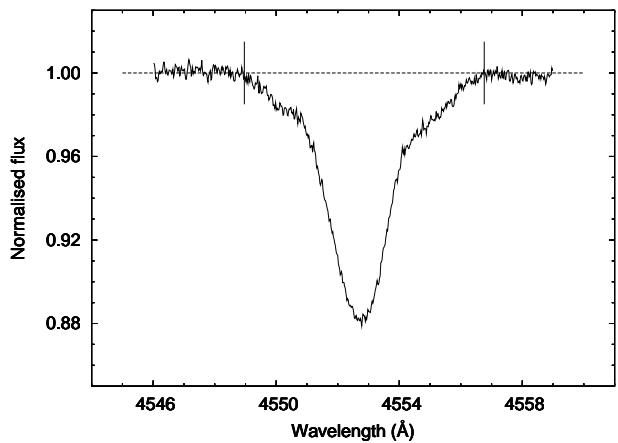}}}
\caption{SiIII $\lambda$\,4552.6\,\AA\ line profile obtained by
averaging 39 spectra taken during a period of 12 nights (3-14
August 2000). The vertical lines demonstrate the estimate of the
total line width of the broad-lined component.}
\label{fig_cen_first_breedte_c1}
\end{figure}
The orbital RVs for the broad-lined component could not be derived
in this way. The following strategy was therefore followed in
Paper\,I.  The full width of the line was derived from epochs when
the radial velocities of both components did not differ much (see
Fig.\,\ref{fig_cen_first_breedte_c1}). It was assumed that this
width is constant, which is a reasonable approach because
rotational broadening is dominant for this component.  The centre
of the broad line, obtained from averaging spectra over a night
(CAT) or over two weeks (Euler), determined by starting from
either its left or its right wing, was taken as a good estimate of
the RV of the broad-lined component.

We show below that these procedures lead to an underestimation of the true RV
values, particularly for the narrow-lined component.  We provide the final RV
values of both components in Table\,1 
{(only available electronically from CDS)}.
Hereafter, we will refer to the star producing the deeper and
narrower SiIII lines as the secondary and to the other component
as the primary. Although this is the opposite of what has been
done in the literature so far, we show that the component with the
broader lines is indeed the more massive of the two.
\addtocounter{table}{1}

\subsection{{\sc korel} disentangling}
\label{cen_korel}

Starting from the orbital solution presented in Paper\,II, we applied {\sc
korel} spectral disentangling (Hadrava 1995, 1997, 2001, 2004b). Although {\sc
korel} was not developed to treat line-profile variations due to oscillations,
Harmanec et al.\ (2005) showed that the code is able to treat such a complex
combination of variability.  

{\sc korel} was applied to our $\beta\,$Cen CES spectra for many different sets
of code input parameters and weights, the CORALIE spectra being too noisy to
allow convergence. The resulting disentangled profiles were evaluated each time
by visual inspection, paying attention to smoothness, symmetry, and the residual
sum of squares.  We considered both the situations where line strengths were
allowed for and were not taken into account.  Moreover, we used several types of
weights. Weights proportional to (S/N)$^2$ turned out to lead to the most stable
solution.  The best results were clearly obtained when {\sc korel} was allowed
to search through a larger subregion in orbital parameter space than indicated
by the uncertainties obtained in Paper\,II.

The adopted {\sc korel} solution was obtained in four subsequent steps, in which
the solution of a particular step was used as the initial guess in the next
step. First the light intensities were kept fixed, next the intensity of the
primary was allowed to vary, then the intensity of the secondary could vary, and
finally the intensities of both components were allowed to vary.
This led to a final {\sc korel} orbital
solution which was slightly, but significantly, different from the
one in Paper\,II, the largest discrepancy occurring in the value
of the semi-amplitude of the secondary's orbit $K_2$: 63.8 $\pm$
0.6 km/\,s (Paper\,II) versus 68.1 km/\,s ({\sc korel}).  The
present version of {\sc korel} unfortunately does not provide
errors in the orbital parameter values given as output.

The dotted lines in Fig.\,\ref{fig_cen_syst_error} show the disentangled
profiles of the primary and secondary shifted to the orbital RV obtained in
Paper\,I.  It is obvious from Fig.\,\ref{fig_cen_syst_error} that the RV
estimate of the secondary derived in Paper\,I is an underestimation of the true
RV and that we must take this into account in the derivation of the physical
parameters of the components.

{In principle, a single application of {\sc korel} disentangling to the
blended line profiles should be enough to obtain the final orbital solution.
In practice, however, tiny changes in the {\sc korel} input
parameters changed the final orbital solution considerably while producing
only small changes in the mean disentangled profiles and rms value. For
this reason, we devised the iteration schemes discussed below.}

\subsection{Preliminary update of the orbital solution}
\begin{figure}
\centering
\rotatebox{-90}{\resizebox{6.0cm}{!}{\includegraphics{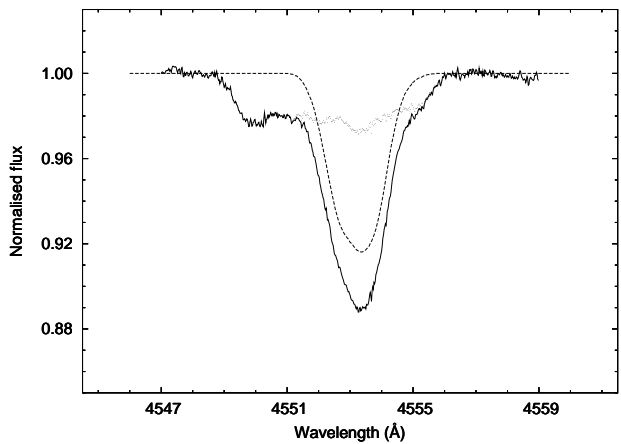}}}
\caption{A SiIII $\lambda$\,4552.6\,\AA\ line observed on 15 March
1998 (full line) and the secondary's disentangled line profile,
which is shifted according to its corresponding orbital RV (dashed
line). The residual spectrum obtained by subtracting the shifted
disentangled line profile from the original spectrum is shown as
dots.}\label{fig_cen_voorbeeld_rm_LPcomp2}
\end{figure}

The disentangled line profile of the secondary allows us to improve the orbital
solution of the primary. We carried out the following procedure for each of the
spectra. We shifted the secondary's disentangled profile according to its
orbital RV obtained in Paper\,I. We subsequently subtracted this shifted
disentangled profile from the original blended line profile.  
{ 
This procedure is
illustrated in Fig.\,\ref{fig_cen_voorbeeld_rm_LPcomp2} for one measurement.
In this way, we
obtain a $\lambda$\,4552.6\,\AA\ line
profile of the primary
whose position and shape is due to the orbital
velocity, as well as to the 
variability due to the oscillations (the dotted line in 
Fig.\,\ref{fig_cen_voorbeeld_rm_LPcomp2}).  
This procedure
resulted in 402 line profiles $(\lambda, I(\lambda))$
which were used to calculate the {\it true\/} RV of the
primary. In view of the bumpy profiles, we cannot use a Gaussian fit to 
compute this true RV. We determine it as follows:
\begin{equation}\label{eq_cen_first_mom}
v_{\rm rad} = \frac{\int (1-I(\lambda))v_\lambda
\,\mathrm{d}\lambda} {\rm EW},
\end{equation}
with {\rm EW} denoting the equivalent width of the line profile
\begin{equation}\label{eq_cen_EW}
{\rm EW} = \int (1-I(\lambda))\,\mathrm{d}\lambda,
\end{equation}
and $v_\lambda$ 
derived by the equation
$$v_\lambda = \frac{\lambda-\lambda_0}{\lambda_0}\,c,$$ with $c$ and $\lambda_0$
the speed of light and the laboratory wavelength, respectively. What we call the
true RV is thus the centroid of the line profile (also termed the first moment,
see Aerts et al.\ 1992 for further extensive discussion of this quantity). Its
computation was done by fixing the integration limits interactively for each
separate profile, after subtracting the disentangled profile of the secondary.}
In this way, 402 RV values of the primary were obtained (compared with only 27
values used in Paper\,I obtained with the procedure described above by means of
Fig.\,2).

Next,
these 402 RV values of the primary were added to the 402 values
for the secondary derived in Paper\,I, and the code {\sc fotel}
(Hadrava 1990, 2004a) was applied to this combined RV dataset.
The results are quite similar to the ones listed in Paper\,II, except that the
semi-amplitude of the primary's orbit, $K_1$, is significantly smaller. This is
consistent with the {\sc korel} disentangling analysis, which led to a higher
value of the semi-amplitude of the secondary, $K_2$ (compared to the value found
in Paper\,II).  Both these results, i.e., a lower $K_1$ value from {\sc korel}
and a higher $K_2$ value after disentangling the secondary's profile,
suggest that the mass ratio $M_1/M_2$ has been underestimated in Paper\,II.

\begin{figure}
\centering
\rotatebox{-90}{\resizebox{6.0cm}{!}{\includegraphics{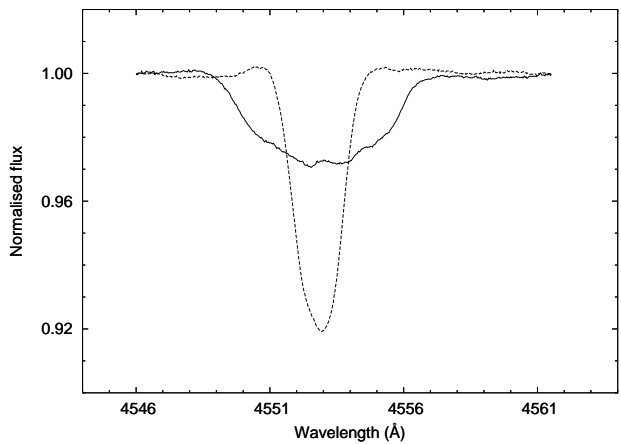}}}
\caption{The disentangled SiIII $\lambda$\,4552.6\,\AA\ line
profiles of the primary (full line) and secondary (dashed
line).}\label{fig_cen_best_dis_profielen}
\end{figure}

Subsequently, a second attempt was undertaken to disentangle the
spectra with {\sc korel} by searching through a neighbouring
subregion of the orbital parameter space centred around the
updated orbital solution with a higher mass ratio.  Again,
satisfactory results were obtained similar to the orbital solution
from {\sc fotel}, except for the values of the semi-amplitudes
$K_1$ and $K_2$.  In comparison with our first attempt to apply
{\sc korel}, the residual sum of squares was lower.
Figure\,\ref{fig_cen_best_dis_profielen} shows the best disentangled
SiIII $\lambda$ 4552.6\,\AA\ line profiles of both components { at this stage
  of the process}.

\subsection{Analysis of the systematic error}

We return to Fig.\,\ref{fig_cen_syst_error}.  The merged line
profile, $I(\lambda)$, is obviously the sum of the primary's
average line profile $I_1(\lambda)$ and the secondary's average
line profile $I_2(\lambda)$. The primary's disentangled
line profile delivers a good approximation of
$I_1$, while the one of the secondary results in an approximation of
$I_2$ (dotted lines in Fig.\,\ref{fig_cen_syst_error}).
The 402 RV values for the secondary 
derived in Paper\,I were obtained by calculating the centroid
of the composite Si III $\lambda$ 4552.6\,\AA\ line profile, i.e., from
Eq.\,(\ref{eq_cen_first_mom}) with $I(\lambda)=I_1(\lambda)+I_2(\lambda)$.  To
{minimize} the interfering influence of the primary's line profile
$I_1(\lambda)$, the integration limits in Eqs.\,(\ref{eq_cen_first_mom}) and
(\ref{eq_cen_EW}) were fixed corresponding to a flux value of 0.95 
{ in Paper\,I,
as indicated in Fig.\,\ref{fig_cen_syst_error}.}
The secondary's RV is, however, given by the centroid of only the secondary's
line profile $I_2(\lambda)$. So, calculating the secondary's RVs 
as was done in Paper\,I, gives each $v_\lambda$ in
the integral (\ref{eq_cen_first_mom})
too high a weight.
When both components have a similar RV, the method adopted in Paper\,I
provides a reasonable approximation of the secondary's RV 
for the following reasons:
\begin{enumerate}
\item the depth of the secondary's line profile has a significantly larger
contribution to the depth of the composite line profile than the depth of the
primary's line profile;
\item the depth of the primary's line profile does not vary much within the
integration limits because of the large width of the primary's line profile
on the one hand and the fact that both components have a similar radial
velocity on the other hand;
\item the equivalent width in the denominator of Eq.\,(\ref{eq_cen_first_mom})
normalises the weighted integral in the numerator.
\end{enumerate}

A systematic error is, however, introduced when the components'
RVs are significantly different. This is the case in
Fig.\,\ref{fig_cen_syst_error}, as the spectrum shown is observed
close to a time at which the primary (secondary) has its minimum
(maximum) orbital RV.  Indeed, in such cases, the second condition
above is not fulfilled, as the depth of the primary's line profile
changes considerably within the integration limits. One can derive
from Fig.\,\ref{fig_cen_syst_error} that, within the integration
limits, the depth of the primary's line profile decreases
monotonically with increasing wavelength so that the additional
weight given to each $v_\lambda$ in Eq.\,(\ref{eq_cen_first_mom})
strongly varies due to the blending of both line profiles. As
lower $v_\lambda$ values systematically get more weight than
higher $v_\lambda$ values, the procedure applied so far still
underestimates the RV of the secondary. We cannot but conclude
that a systematic error has been introduced in Paper\,I in the
calculation of the RVs, due to the fact that the line profiles of
both components are so strongly blended with each other.

\subsection{Iterative determination of the orbital parameters}
\label{cen_iter}

Since the orbital parameters common to the fit to both the spectroscopic and
interferometric data were in agreement with each other, it is appropriate to
assume that this spectroscopically determined orbital solution is already close
to the true orbit and can hence be considered as a good initial solution to
start an iterative process to improve the orbital parameter values. Two
different iteration schemes were applied.

\subsubsection{Iteration scheme I}
\label{cen_iter_1}
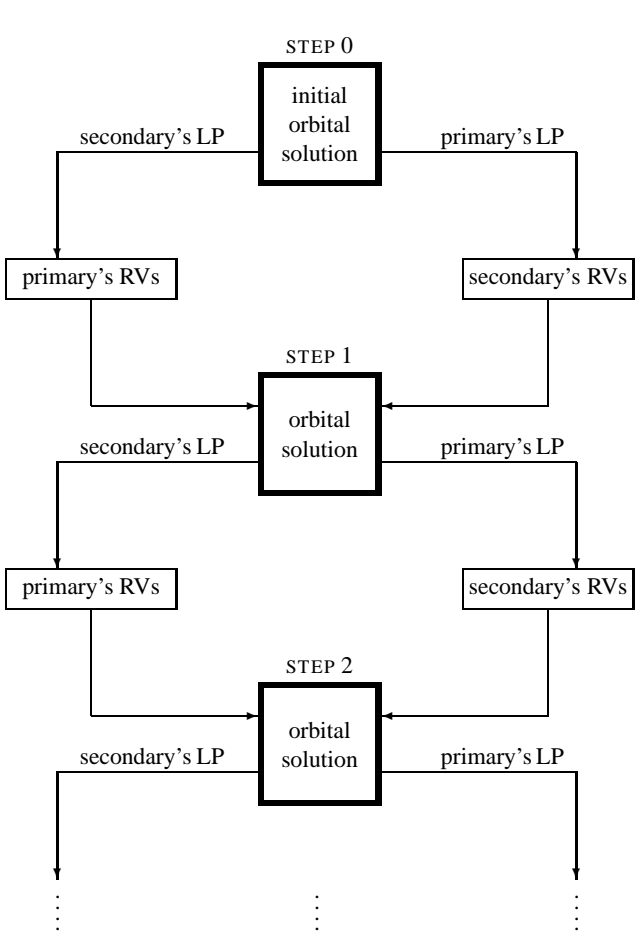
\begin{figure}
\centering
\setlength{\unitlength}{0.75mm}
\begin{picture}(113,170)
\label{diag_cen_schemeI}

\linethickness{0.7mm}

 \multiput(46.5,30)(0,55){2}{\framebox(20,20){\begin{minipage}{20mm}
\centerline{orbital} \centerline{solution}\end{minipage}}}
 \put(46.5,140){\framebox(20,20){\begin{minipage}{20mm}\centerline{initial}
\centerline{orbital} \centerline{solution}\end{minipage}}}

\thinlines

 \put(46.5,161){\makebox(20,6){\sc step 0}}
 \put(46.5,106){\makebox(20,6){\sc step 1}}
 \put(46.5,51){\makebox(20,6){\sc step 2}}

 \put(55.5,7){\shortstack{.\\.\\.\\.}}
 \put(14,146.5){secondary's\,LP}
 \put(45.4,145){\line(-1,0){35.5}}
 \put(10,145){\vector(0,-1){19}}
 \put(1,119){\framebox(30,7){primary's RVs}}
 \put(16,119){\line(0,-1){19}}
 \put(16,100){\vector(1,0){29.5}}

 \put(14,91.5){secondary's\,LP}
 \put(45.4,90){\line(-1,0){35.5}}
 \put(10,90){\vector(0,-1){19}}
 \put(1,64){\framebox(30,7){primary's RVs}}
 \put(16,64){\line(0,-1){19}}
 \put(16,45){\vector(1,0){29.5}}

 \put(14,36.5){secondary's\,LP}
 \put(45.4,35){\line(-1,0){35.5}}
 \put(10,35){\vector(0,-1){19}}
 \put(9.5,7){\shortstack{.\\.\\.\\.}}
 \put(78,146.5){primary's\,LP}
 \put(66.5,145){\line(1,0){35.5}}
 \put(102,145){\vector(0,-1){19}}
 \put(82,119){\framebox(30,7){secondary's RVs}}
 \put(97,119){\line(0,-1){19}}
 \put(97,100){\vector(-1,0){29.5}}

 \put(78,91.5){primary's\,LP}
 \put(66.5,90){\line(1,0){35.5}}
 \put(102,90){\vector(0,-1){19}}
 \put(82,64){\framebox(30,7){secondary's RVs}}
 \put(97,64){\line(0,-1){19}}
 \put(97,45){\vector(-1,0){29.5}}

 \put(78,36.5){primary's\,LP}
 \put(66.5,35){\line(1,0){35.5}}
 \put(102,35){\vector(0,-1){19}}
 \put(101.5,7){\shortstack{.\\.\\.\\.}}
\end{picture}
\caption{Flowchart diagram of Iteration Scheme I (see text for details).}
\end{figure}
Figure\,5
summarises the features of the first iteration scheme. In each
iteration step, a new orbital solution is calculated by applying
{\sc fotel} to the combined dataset of new RV values of both the
primary and secondary. The new set of the primary's RVs is obtained by
calculating the centroid of the ``secondary subtracted line
profiles''. The latter refer to the spectra obtained by taking the
difference between the original spectra and the shifted
secondary's disentangled line profile (abbreviated as
``secondary's LP'' in Fig.\,4). The shift corresponds to the
orbital velocity value given by the orbital solution of the
previous iteration step. This procedure to determine the secondary
subtracted line profiles is illustrated in
Fig.\,\ref{fig_cen_voorbeeld_rm_LPcomp2}. A new set of the 
secondary's
RV values is calculated in each iteration step in an analogous
manner by means of the primary's disentangled line profile
(abbreviated as ``primary's LP'' in Fig.\,4). The primary's and
secondary's disentangled line profiles mentioned above are
properly normalised versions of the best disentangled profiles
obtained with {\sc korel}.
All centroid velocity values were made with fixed integration
limits to reduce the noise level. The widths of the primary's and
secondary's disentangled line profiles provide us with an
objective way to select these fixed integration limits.

\subsubsection{Iteration scheme II}
\label{cen_iter_2}

The second iteration scheme is similar to the first one, but
slightly more complicated. For the sake of clarity, the reader is
advised to refer to Fig.\,6
while reading the following description.

\begin{figure}
\centering
\setlength{\unitlength}{0.75mm}

\begin{picture}(113,170)
\label{diag_cen_schemeII}

\linethickness{0.7mm}

 \multiput(45.5,10)(0,45){3}{\framebox(20,20){\begin{minipage}{20mm}
\centerline{orbital} \centerline{solution}\end{minipage}}}
 \put(45.5,145){\framebox(20,20){\begin{minipage}{20mm}\centerline{initial}
\centerline{orbital} \centerline{solution}\end{minipage}}}

\thinlines

 \put(45.5,167){\makebox(20,6){\sc step 0}}
 \put(45.5,121){\makebox(20,6){\sc step 1}}
 \put(45.5,76){\makebox(20,6){\sc step 2}}
 \put(45.5,32){\makebox(20,6){\sc step 3}}

 \put(55.5,0){\shortstack{.\\.\\.\\.}}


 \put(14,157){secondary's\,LP}
 \put(45.5,155){\line(-1,0){35.5}}
 \put(10,155){\vector(0,-1){23}}

 \put(1,125){\framebox(30,7){primary's RVs}}
 \put(26,125){\line(0,-1){15}}
 \put(26,110){\vector(1,0){18.9}}

 \put(11,125){\vector(0,-1){5}}
 \put(1,100){\framebox(20,20){\begin{minipage}{20mm}\centerline{orbital}
\centerline{solution}\centerline{primary}\end{minipage}}}
 \put(11,100){\vector(0,-1){13}}
 \put(14,92){primary's\,LP}

 \put(1,80){\framebox(30,7){secondary's RVs}}
 \put(26,80){\line(0,-1){15}}
 \put(26,65){\vector(1,0){18.9}}

 \put(11,80){\vector(0,-1){5}}
 \put(1,55){\framebox(20,20){\begin{minipage}{20mm}\centerline{orbital}
\centerline{solution}\centerline{secondary}\end{minipage}}}
 \put(11,55){\vector(0,-1){13}}
 \put(14,47){secondary's\,LP}

 \put(1,35){\framebox(30,7){primary's RVs}}
 \put(26,35){\line(0,-1){15}}
 \put(26,20){\vector(1,0){18.9}}

 \put(11,35){\vector(0,-1){5}}
 \put(10.5,21){\shortstack{.\\.\\.\\.}}


 \put(77,157){primary's\,LP}
 \put(66.5,155){\line(1,0){35.5}}
 \put(102,155){\vector(0,-1){23}}

 \put(82,125){\framebox(30,7){secondary's RVs}}
 \put(87,125){\line(0,-1){15}}
 \put(87,110){\vector(-1,0){20.5}}

 \put(102,125){\vector(0,-1){5}}
 \put(92,100){\framebox(20,20){\begin{minipage}{20mm}\centerline{orbital}
\centerline{solution}\centerline{secondary}\end{minipage}}}
 \put(102,100){\vector(0,-1){13}}
 \put(75,92){secondary's\,LP}

 \put(82,80){\framebox(30,7){primary's RVs}}
 \put(87,80){\line(0,-1){15}}
 \put(87,65){\vector(-1,0){20.5}}

 \put(102,80){\vector(0,-1){5}}
 \put(92,55){\framebox(20,20){\begin{minipage}{20mm}\centerline{orbital}
\centerline{solution}\centerline{primary}\end{minipage}}}
 \put(102,55){\vector(0,-1){13}}
 \put(77,47){primary's\,LP}

 \put(82,35){\framebox(30,7){secondary's RVs}}
 \put(87,35){\line(0,-1){15}}
 \put(87,20){\vector(-1,0){20.5}}

 \put(102,35){\vector(0,-1){5}}
 \put(101.5,21){\shortstack{.\\.\\.\\.}}
\end{picture}
\caption{Flowchart diagram of Iteration Scheme II (see text for details).}

\end{figure}
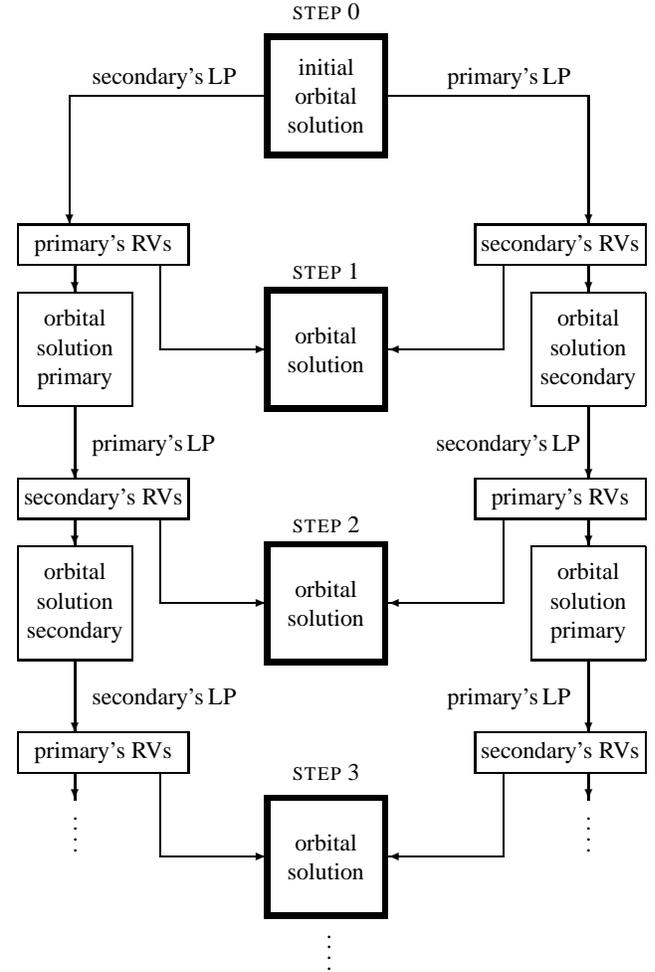

In each iteration step, a new orbital solution is again derived on
the basis of the combined dataset that consists of the newly
derived RV values of both components. Just as in iteration scheme
I, the new primary's (secondary's) RVs are derived by calculating
the centroid of the ``secondary's (primary's) subtracted line
profiles'' with fixed integration limits according to the width of
the disentangled line profile of either component. The difference
between the two iteration schemes lies in the way that these
``primary's/secondary's subtracted line profiles'' are
constructed. For each original spectrum, both schemes shift the
primary's (secondary's) disentangled line profile according to the
corresponding orbital velocity that was found in the previous
iteration step and subtract it from the original spectrum. While
iteration scheme I uses the orbital RV corresponding to the
orbital solution derived on the basis of the old RV datasets of
both components, iteration scheme II derives the new primary's
(secondary's) RV dataset by using the orbital solution derived on
the basis of the old secondary's (primary's) RVs.

\begin{figure}
\centering
\resizebox{8.0cm}{!}{\includegraphics{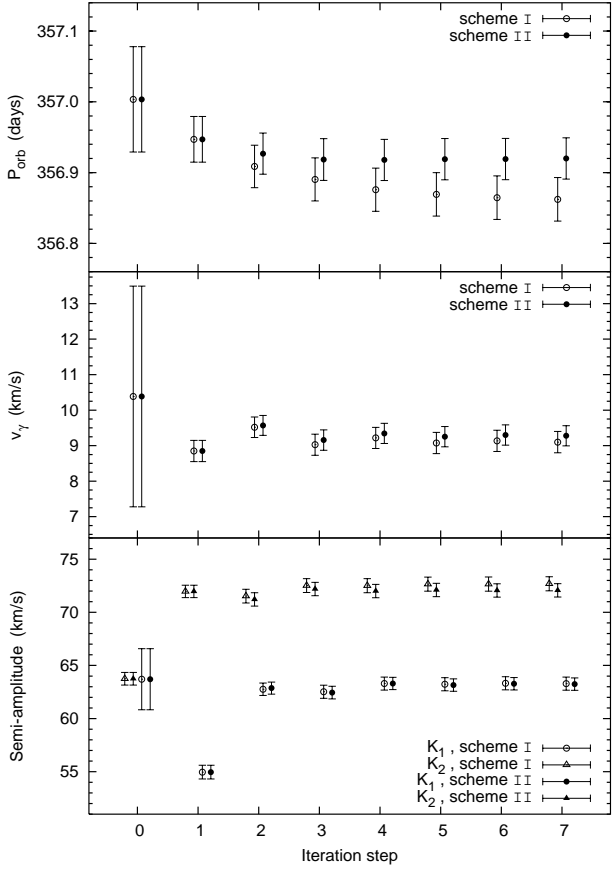}}
\caption{Comparison of the results of iteration scheme I (open
symbols) and II (filled symbols): evolution of the orbital period
$P_{\rm orb}$ [upper panel], system velocity $v_\gamma$ [middle
panel], and the semi-amplitudes $K_1$ (dots) and $K_2$ (triangles)
[lower panel]. The vertical bars denote the errors provided by
{\sc fotel}.} \label{fig_cen_3_1_tot_cross_vast}
\end{figure}
\begin{figure}
\centering
\resizebox{8.0cm}{!}{\includegraphics{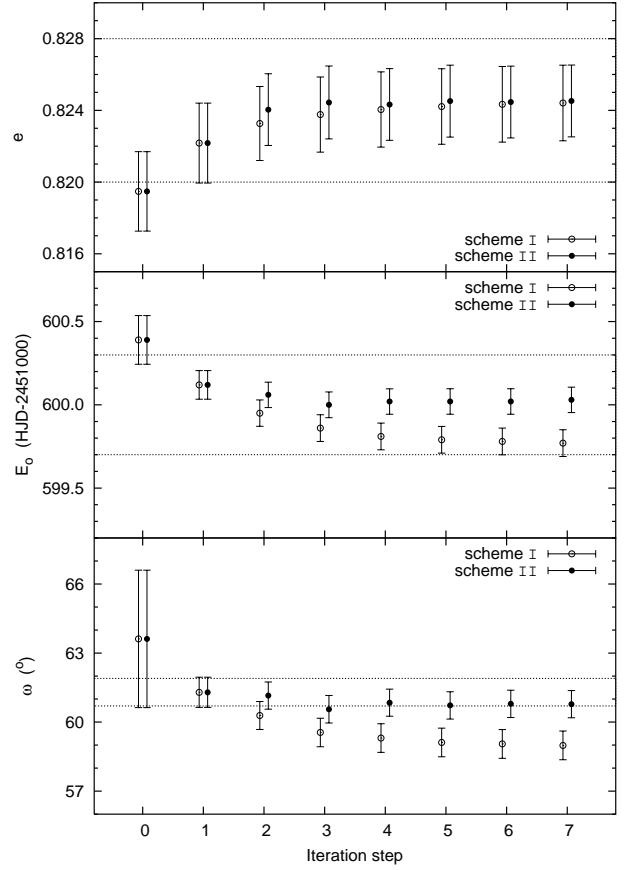}}
\caption{Comparison of the results of iteration scheme I (open
symbols) and II (filled symbols): evolution of the eccentricity
$e$ [upper panel], epoch of periastron passage $E_{\rm o}$ [middle
panel], and the longitude of periastron $\omega$ [lower panel]. The
two horizontal, dotted lines in each panel indicate the
corresponding interval estimated from the interferometric
measurements taken from Paper\,II.  The vertical bars denote the
errors provided by {\sc fotel}.}
\label{fig_cen_3_2_tot_cross_vast}
\end{figure}

Figures\,\ref{fig_cen_3_1_tot_cross_vast} and \ref{fig_cen_3_2_tot_cross_vast}
show the results of the application of iteration schemes I (open symbols) and II
(filled symbols).  It reveals only minor differences in the final parameter
values obtained with both schemes. Scheme II leads to a slightly higher value of
the orbital period, but the differences in the system velocity as well as the
semi-amplitudes of both components' orbit are totally negligible.  The stability
of the semi-amplitudes $K_1$ and $K_2$ is of particular importance as these
parameters allow an accurate mass determination of both components.  The upper
panel of Fig.\,\ref{fig_cen_3_2_tot_cross_vast} shows that both iteration
schemes make the eccentricity converge to a value in excellent
agreement with the interferometric value. One can derive from the middle panel
of Fig.\,\ref{fig_cen_3_2_tot_cross_vast} that iteration scheme II puts the
value of the epoch of periastron passage closer to the centre of the interval
which was derived for this parameter on the basis of interferometric data than
scheme I.  The lower panel of Fig.\,\ref{fig_cen_3_2_tot_cross_vast} reveals an
increase in the value of the longitude of periastron which eliminates the small
discrepancy which appeared when applying iteration scheme I.

\subsubsection{Evaluation of both iteration schemes}

The final orbital solutions derived with iteration schemes I and
II are listed in  the second and third columns of
Table\,\ref{tab_cen_iterative_orbit}, respectively.

\begin{table}
\tabcolsep=2pt
\centering
\caption{Orbital parameters for $\beta$\,Cen obtained from {\sc korel}
  disentangling and by the
application of iteration processes to the observed SiIII $\lambda$ 4552.6\,\AA\
line profiles listed in Paper\,I. The errors are 1$\sigma$ estimates resulting
from {\sc fotel} assuming the {\sc korel} disentangled profiles to be
error-free. }
\begin{tabular}{llll}
\hline
{Parameter} & {\sc korel} & {Scheme I} & {Scheme II} \\
\hline
$P_{\rm orb}$ (days) & 356.94 & 356.86 $\pm$ 0.03 & 356.92 $\pm$ 0.03 \\[0mm]
$v_{\gamma}$ (km\,s$^{-1}$) & -- & 9.1 $\pm$ 0.3 & 9.3 $\pm$ 0.3 \\[0mm]
$K_1$ (km\,s$^{-1}$) & 57.4 & 63.3 $\pm$ 0.6 & 63.2 $\pm$ 0.6\\[0mm]
$K_2$ (km\,s$^{-1}$) & 72.3 & 72.7 $\pm$ 0.7 & 72.1 $\pm$ 0.6 \\[0mm]
$e$  & 0.825 & 0.824 $\pm$ 0.002 & 0.825 $\pm$ 0.002 \\[0mm]
$E_0$ (HJD) & 2451600.08 & 2451599.77 $\pm$ 0.08 & 2451600.03 $\pm$ 0.08 \\[0mm]
$\omega$ ($^{\rm o}$) & 62.2 & 59.0 $\pm$ 0.6 & 60.8 $\pm$ 0.6 \\[0mm]\hline
\end{tabular}
\label{tab_cen_iterative_orbit}
\end{table}

To reveal the origin of the small differences between the
results obtained with the two iteration schemes, we examined the
dataset consisting of the differences between the final RV values
derived with iteration schemes II and I:
$$\{\Delta v(t)\}_t = \{v_{\rm rad,\,II}(t) - v_{\rm rad,\,I}(t)\}_t.$$
This dataset includes two subsets: the primary's and secondary's
radial velocity differences between schemes II and I:
$$\{\Delta v(t)\}_t = \{\Delta v_{\rm\,primary}(t)\}_t \cup \{\Delta v_{\rm\,
secondary}(t)\}_t.$$ For each of the above three datasets, the average and the
standard deviation were calculated. It is clear from the results, which are
listed in Table\,\ref{tab_cen_mean_sd}, that the secondary's RVs
are in much better agreement than those of the primary. The larger
average difference of the $\{\Delta v_{\rm\,primary}(t)\}_t$ dataset is due to
the primary's line profile being so strongly rotationally broadened.  This
implies that 
the orbital solution derived by means of only the secondary's final RVs
is very stable regardless of the applied iteration scheme, while the orbital
solution derived on the basis of only the primary's final RVs is somewhat less
stable and, hence, causes the small difference between the final results of
iteration scheme I and II. We consider the results obtained with iteration
scheme II more reliable because they are in better agreement with the
interferometrically derived orbital solution on the one hand and, although it is
not clear in the rounded values listed in Table\,\ref{tab_cen_iterative_orbit},
the uncertainties in the derived parameters derived with scheme II are all
smaller.

We also note that both iteration schemes lead to smaller error bars on the
derived orbital parameter values than the ones obtained in
Paper\,I. Figures\,\ref{fig_cen_3_1_tot_cross_vast} and
\ref{fig_cen_3_2_tot_cross_vast} show that the uncertainties in the system
velocity, the semi-amplitude of the primary's orbit, and the longitude of
periastron have been significantly lowered by the iterative process. The
application of iteration scheme II maintains or even improves the compatibility
with the interferometric results (see Fig.\
\ref{fig_cen_3_2_tot_cross_vast}). Therefore, we conclude that the iterative
process results in a significant improvement of the orbital solution.

\begin{figure*}
\centering \resizebox{.9\textwidth}{!}{\rotatebox{-90}
{\includegraphics{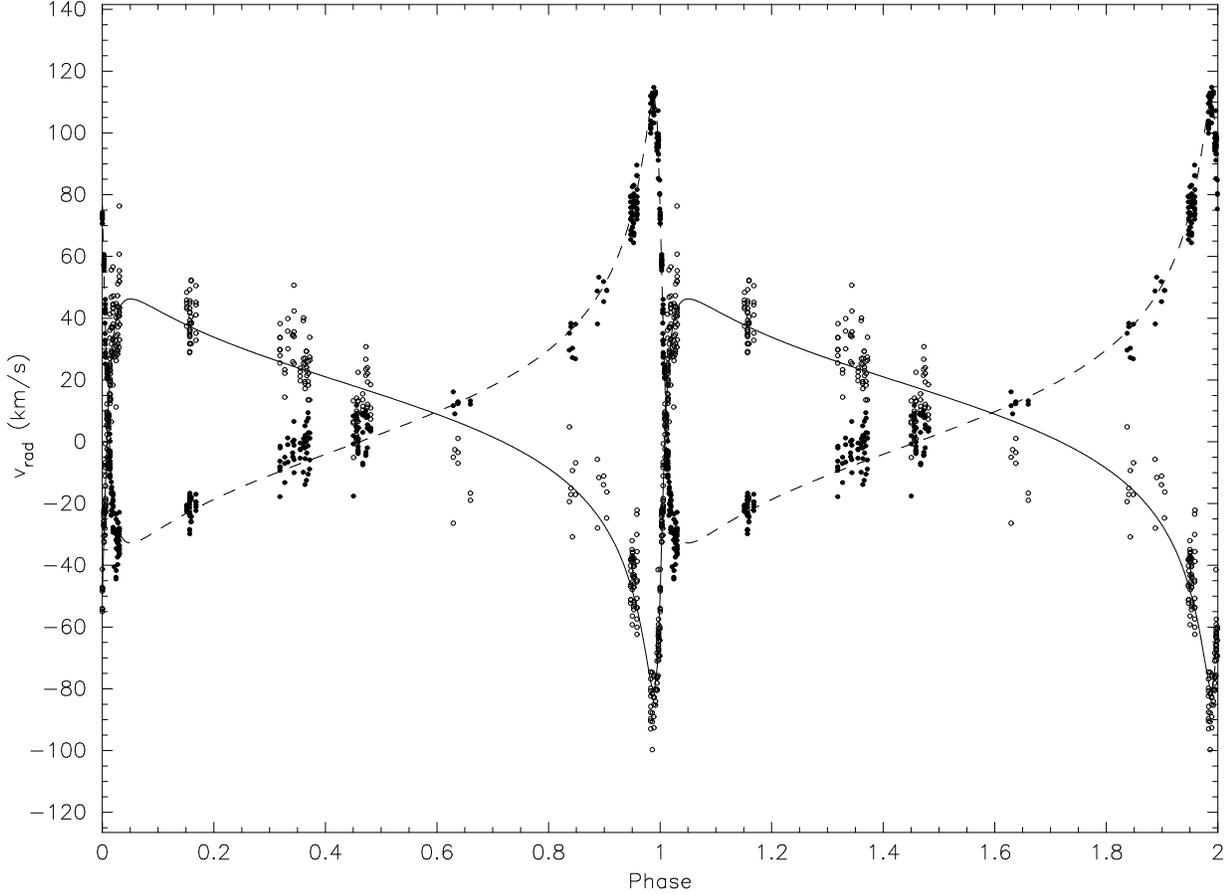}}} \caption{The orbital curves
obtained by means of iteration scheme II. The open and filled dots
represent the RVs of, respectively, the primary and secondary
components around the centre of mass.  The full and dashed lines
represent the best-fitting orbits of the primary and
secondary, respectively, according to the parameters listed in the column
labelled ``Scheme II'' in
Table\,\protect\ref{tab_cen_iterative_orbit}.}
\label{fig_cen_final_baan}
\end{figure*}

\begin{table}
\centering
\tabcolsep=2pt
\caption{Statistical properties of the differences between the final RV values
derived with iteration scheme II and I. See text for more information on the
three datasets.}
\begin{tabular}{ccc}\hline
Dataset & Average (km\,s$^{-1}$) & Standard deviation (km\,s$^{-1}$)\\
\hline
$\{\Delta v_{\rm\,primary}(t)\}_t$ & 2.37 & 1.02 \\
$\{\Delta v_{\rm\, secondary}(t)\}_t$ & 0.26 & 0.27 \\
$\{\Delta v(t)\}_t$ & 1.32 & 0.75 \\\hline
\end{tabular}\label{tab_cen_mean_sd}
\end{table}
The final primary's (open dots) and secondary's (filled dots) RV values that
were obtained with iteration scheme II are shown in
Fig.\,\ref{fig_cen_final_baan}. The best combined fit to these datasets is
denoted as a full (primary's orbit) or dashed (secondary's orbit) line. The fit
is satisfying compared to the one obtained in Paper\,I.

Finally, we applied {\sc korel} again with this new, iteratively derived orbital
solution. Indeed, in theory this can result in improved versions of the
disentangled profiles of both components and, therefore, allow an iteration
process on a higher level.  However, the best disentangled profiles that came
out of the {\sc korel} analysis were hardly distinguishable from the ones shown
in Fig.\,\ref{fig_cen_best_dis_profielen} and, hence, there is no point in
repeating the iteration process with these ``new'' disentangled profiles.  
{
In particular, the small bump in the centre of the primary's disentangled
profile in
Fig.\,\ref{fig_cen_best_dis_profielen} did not disappear. It is due to the
imperfect averaging over the oscillations of the primary by {\sc korel}. This is
not surprising in view of its complex multiperiodic high-degree oscillations,
which we cannot unravel perfectly from our data (see Sect.\,4).}

\section{Physical parameters of the components}
\label{cen_glob_par}

The physical parameters of $\beta\,$Cen derived in Paper\,II were
obtained without taking into account the systematic effects
described here. Their values and errors must clearly be revised.
The error estimates were optimistic as they were derived from
systematically underestimated RV values.

We first checked that the refined value of the orbital period
(which was fixed in deriving the interferometric orbital
parameters in Paper\,II) does not change the values of the orbital
inclination and angular semi-major axis.  We then followed the
same strategy for the computation of the individual component
masses as in Paper\,II, using the refined values of the orbital
period and the semi-amplitudes of both components from scheme II,
the orbital inclination from the interferometry given in
Paper\,II, and the mean of the spectroscopic and interferometric
values for the eccentricity. We also need to take into account the
standard errors of these quantities. This is not so
straightforward for the parameters resulting from our iteration
schemes because their 1$\sigma$ error from {\sc fotel} listed in
Table\,\ref{tab_cen_iterative_orbit} is necessarily an
underestimation of the true error. Indeed, these errors were
derived under the assumption that the disentangled profiles
resulting from {\sc korel} are error-free, which is not the case.
We are, unfortunately, unable to propagate the uncertainty induced
by the disentangling properly because the current version of {\sc
korel} does not provide us with error estimates. Moreover, our
schemes implicitly assume that {\sc korel} appropriately treats
the effects of the oscillations as random noise in computing the
disentangled profiles. For this reason, we adopt a very
conservative approach for the error propagation and use 2$\sigma$
errors rather than those listed in
Table\,\ref{tab_cen_iterative_orbit} in the derivation of the
physical parameters of the system. Following the approach of
Paper\,II, this leads to $M_1 = 11.2 \pm 0.7 \,M_\odot$ and $M_2 =
9.8 \pm 0.7\,M_\odot$.  Finally, this explains our choice for the
revised nomenclature of the primary and secondary.

We attempted to refine these estimates by using all the additional
observational information we have at our disposal, such as the
CORALIE \'echelle spectra.  We first estimated the effective
temperature and gravity of the two components by following the
procedure outlined in Uytterhoeven et al.\ (2005) for the
double-lined binary $\kappa\,$Sco, i.e., by merging theoretical
line profiles of H, He, and Si lines with the appropriate flux ratio
according to NLTE predictions made from the latest version of the
FASTWIND code (Puls et al.\ 2005), after using the orbital RVs to
shift the profiles. This led us to the conclusion that both
components have $T_{\rm eff}=24\,000\pm 1\,000$K and
$\log\,g=3.4\pm 0.3$. The large uncertainty in the gravity stems
from the difficulty in achieving a proper {normalisation} of the
spectra near the Balmer lines. Since we find the two components to
have equal $T_{\rm eff}$ and $\log\,g$ within the uncertainties,
it is possible to compute photometric estimates of these
quantities from multicolour photometry. We did this from Geneva
measurements of $\beta\,$Cen at our disposal and find $T_{\rm
eff}=26\,500\pm 500$K and $\log\,g=3.7\pm 0.2$ assuming equal
components. This leads us to a safe broad range of $T_{\rm
eff}=25\,000\pm 2\,000$K and $\log\,g=3.5\pm 0.4$ for both
components.

We subsequently scanned the very extensive database of
main-sequence stellar models published by Ausseloos et al.\
(2004), which have a range in mass from 7 to 13$\,M_\odot$ in
steps of $0.1\,M_\odot$ and a range in $Z$ from 0.012 to 0.030 in
steps of 0.002, for each of the three values of the core
overshooting parameter of 0.0, 0.1, and 0.2 expressed in local
pressure scale heights. The models have $X=0.70$ and the solar
mixture of Grevesse et al.\ (1996). For a description of the input
physics, we refer to Ausseloos et al.\ (2004). We scanned this
database requesting that the masses, effective temperatures, and
gravities of $\beta\,$Cen's components lie in the appropriate
ranges and that the age of the components must be equal to within
1\%. This leads us to acceptable ranges for the masses of $M_1\in
[10.6,10.8]\,M_\odot$ and $M_2\in [10.2,10.4]\,M_\odot$, and an
age~$\in [13.5,14.7]$ million years.

As an {\it a posteriori\/} check, we computed the allowed mass ratio resulting
from the brightness ratio obtained from the interferometry in Paper\,II and the
mass-luminosity relation $\log (L_2/L_1)=(3.51\pm 0.14) \log (M_2/M_1)$
(Griffiths et al.\ 1988). We thus find the condition $M_2/M_1\in[0.95,0.97]$,
which is fulfilled by our solutions for the masses resulting from the
spectroscopy and interferometry refined by the modelling.

Finally, the systematic errors in the semi-amplitudes also call for a
re-evaluation of the dynamical parallax given in Paper\,II. Following the same
approach as in Paper\,II and using $2\sigma$ errors for the spectroscopic
elements, we find $\pi = 9.3\pm 0.3$, resulting in a distance of 108$\pm$4 pc.

\section{Analysis of the line-profile variability}

Challenging aspects of massive star asteroseismology are the detection of
numerous frequencies and their mode identification on the one hand, and the
derivation of the fundamental parameters of the targets ($T_{\rm eff}$,
$\log\,g$, $M$) with high precision on the other hand. We succeeded in the
latter, and make an attempt to tackle the former challenge now.  

\subsection{Frequency analysis}

Our aim is to find the timescales associated with the short-term line-profile
variability and to connect each timescale with the component to which it
belongs. This is by no means straightforward because the short-term variations
have a significantly lower amplitude than the orbital variations.  Moreover, the
line profiles of both components are fully blended with each other at all
orbital phases.  {This required a specific non-standard analysis, the details
of which are available in Ausseloos (2005). Here, we present only a concise
summary of the results. In particular, we point out that the complexity of the
profile variations due to the presence of moving subfeatures (see Ausseloos
2005, p. 28, Fig.\,2.1 for examples and
Fig.\,\ref{fig_cen_voorbeeld_rm_LPcomp2}) does not favour a standard
radial-velocity analysis, but requires a search for frequencies across the whole
width of the profiles.}

We perform an analysis of the intrinsic variability of the primary by using a
two-dimensional (2D) frequency analysis method first introduced by Gies \&
Kullavanijaya (1988) and later defined as the Intensity Period Search (IPS) by
Telting et al.\ (1997).  We used a 2D version of the Lomb-Scargle method
(Scargle 1982) for the time series of normalised flux values at each wavelength
across the profile.  As we found a value of 0.0045\,c\,d$^{-1}$ for the
half-width at half-maximum of the central peak of the window function based on
all CES data, and as the 2D frequency analysis is rather time consuming, we
adopted a frequency step of 0.001\,c\,d$^{-1}$ in a first stage. After having
identified the main peak and its aliases, we recomputed the periodograms with a
factor 10 smaller frequency step around the dominant peak and its aliases to
check if the results remained valid, which was always the case for the relevant
frequencies mentioned below.
\begin{figure}
\rotatebox{-90}{\resizebox{6.0cm}{!}{\includegraphics[clip=true,
viewport=0cm 2.5cm 17cm 28cm]{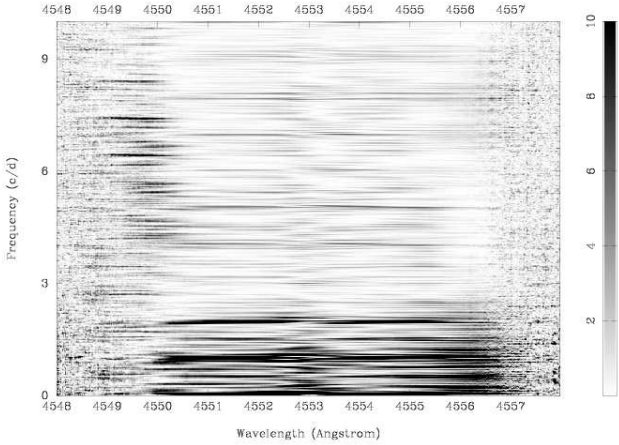}}} 
\caption{Grey-scale representation of the power spectrum obtained at each
wavelength position (given in units of the variance) across the SiIII $\lambda$
4552.6\,\AA\ line profile by the application of the 2D Lomb-Scargle method to
all CES spectra.
}\label{fig_cen_multishow_cat+cat93}
\end{figure}

There exists a small wavelength range centred around 4549\,\AA, at which only
the primary's line profile is present. We first applied the 2D Lomb-Scargle
method to the dataset comprising all the CES spectra. A graphical representation
of the results is shown in Fig.\,\ref{fig_cen_multishow_cat+cat93}. This figure
reveals power excess in the wavelength range between 4550\,\AA\ and 4557\,\AA,
which is visible at frequencies below 3 c\,d$^{-1}$. This is caused by the shift
of the secondary's line profile due to its orbital motion. However, clearly
visible peaks occur between 5 and 9 c\,d$^{-1}$ in the power spectra at
wavelengths between 4549\,\AA\ and 4550\,\AA. This excess power can only be due
to the primary, proving that this component has short-term periodic variability.

\begin{figure}
\centering
\rotatebox{-90}{\resizebox{6.0cm}{!}{\includegraphics[clip=true,
viewport=0cm 2.5cm 17cm 28cm]{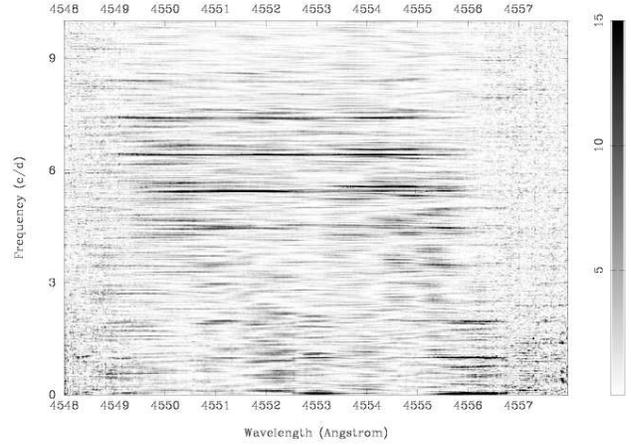}}}
\caption{Same as Fig.\,\protect\ref{fig_cen_multishow_cat+cat93}, but 
after subtracting the secondary's disentangled line profile and correcting
for the primary's orbital motion.}\label{fig_cen_multishow_cat_na_rm_LP}
\end{figure}

To find the frequencies of the short-term variability of the primary
with better significance, we removed the higher-amplitude variability due to its
orbital motion around the centre of mass.  For this, we used the secondary's
disentangled line profile in combination with the orbit. We then computed the
periodograms at each wavelength to construct a 2D periodogram. The results for
the CES spectra are shown in Fig.\,\ref{fig_cen_multishow_cat_na_rm_LP} (we
omitted the CORALIE spectra for this plot due to their larger noise level). We
clearly reveal excess power across the whole line profile of the primary 
at a frequency near 6.4\,c\,d$^{-1}$ and its
aliases.
\begin{figure}
\centering
\resizebox{7.5cm}{!}{\includegraphics{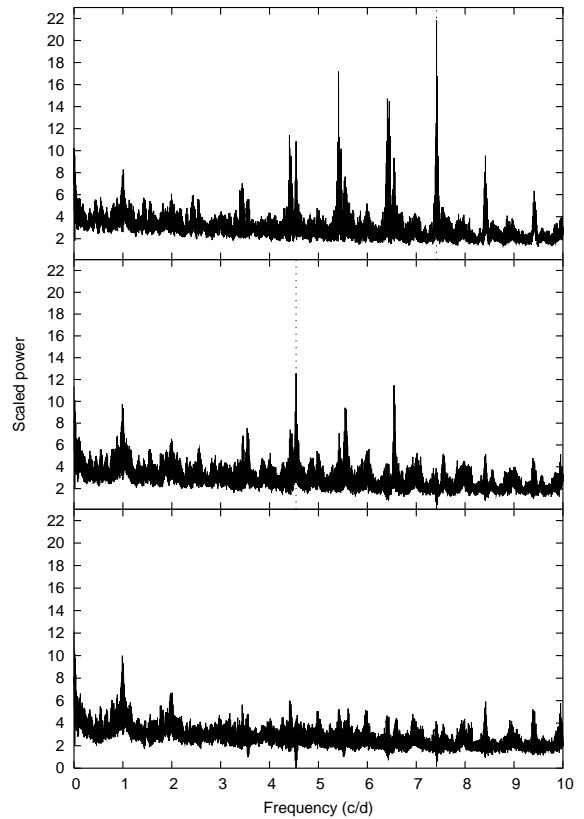}}
\caption{Power spectra resulting from the addition of all the 1D periodograms
over the wavelength range between 4549 and 4556.5\,\AA.  The panels correspond
to different stages of prewhitening: original periodogram (upper panel), after
prewhitening with 7.415\,c\,d$^{-1}$ (middle panel), and after prewhitening with
7.415 and 4.542\,c\,d$^{-1}$ (lower panel).}
\label{fig_cen_power_all_na_rm_LP}
\end{figure}

We summed all the 1D periodograms (for both the CES and CORALIE spectra) over
the range 4549 and 4556.5\,\AA\ (Fig.\,\ref{fig_cen_power_all_na_rm_LP}).  The
drawback is that this analysis method removes the secondary's long-term
variability, but not its short-term variability. Therefore, the sum of all 1D
periodograms shown in Fig.\,\ref{fig_cen_power_all_na_rm_LP} contains a mix of
peaks due to the primary's and secondary's short-term variability, if any.  The
upper panel unquestionably reveals 7.415\,c\,d$^{-1}$ as the dominant frequency,
although rather strong aliasing still occurs for the whole dataset. We have
shown above that this frequency belongs to the primary. The middle
panel of Fig.\,\ref{fig_cen_power_all_na_rm_LP} shows the periodogram that is
obtained after prewhitening the original data with 7.415\,c\,d$^{-1}$. It
suggests 4.542\,c\,d$^{-1}$, or one of its aliases, as the second frequency.
After prewhitening with 7.415 and 4.542\,c\,d$^{-1}$, the power spectrum (lower
panel) is dominated by peaks at lower frequencies. The highest peak in the
interval [3,10]\,c\,d$^{-1}$ appropriate for $\beta\,$Cephei stars occurs at
4.407\,c\,d$^{-1}$, but it is not clear at this stage whether this is another
intrinsic frequency.  {We conclude that there is evidence for
unexplained additional power.}

A study of the secondary's intrinsic temporal behaviour is by no means
straightforward. Due to their large width, the primary's line profiles extend
over a wavelength range that completely includes the range spanned by the
secondary's line profiles during nearly all orbital phases. This makes it very
difficult to unravel the secondary's line-profile variations, if any, from the
ones of the primary.  We applied the following procedure: for each original
spectrum, the primary's SiIII $\lambda$\,4552.6\,\AA\ disentangled line profile
was shifted according to the corresponding orbital velocity and subsequently
subtracted from the original spectrum. We then computed a 2D Scargle periodogram
as explained above.  We subsequently added the power across subintervals of the
total wavelength range [4547, 4557]\,\AA. At the same time, we determined the
extent of the secondary's line within that wavelength range 
(see Ausseloos 2005,
p.72, Fig.\,2.32 for more details).  If the power is only significant in the
wavelength range spanned by the secondary, this is considered as a strong
indication that the corresponding frequency is associated with that star.

We carried out several tests by considering different
subdatasets whose power distribution for candidate frequencies was calculated
over different subintervals in wavelength.  From all these tests, we conclude
that $f_2$ also belongs to the primary and that no evidence of short-term
variability in the line profiles of the secondary was found in our dataset. It
has to be stressed that all previously published period analyses of $\beta$\,Cen
(Breger 1967; Shobbrook \& Robertson 1968; Lomb 1975; Robertson et al.\ 1999;
Ausseloos et al.\ 2002) can no longer be trusted as they all neglect the
pulsations of the primary and assume that the component with the deeper and
narrower line profiles undergoes the short-term variability.

\subsection{Mode identification}

An attempt was made to identify the modes of the primary. Given its strongly
rotationally-broadened profiles, the Doppler Imaging method was used.  This
method was introduced by Gies \& Kullavanijaya (1988) for the B0.7III star
$\varepsilon\,$Per, but several authors have elaborated on it since (see, e.g.,
Telting \& Schrijvers 1997 and references therein).  Telting \& Schrijvers
(1997) took a major step forward by making a large Monte-Carlo simulation study
from which they derived linear relationships between the degree $\ell$ and the
blue-to-red phase difference $\Delta\Psi_f$ of an observed frequency $f$ on the
one hand, and between the azimuthal number $m$ and the phase difference of the
first harmonic $\Delta\Psi_{2f}$ on the other hand.  The errors of the estimates
of $\ell$ and $m$ are, respectively, one and two. Telting \& Schrijvers (1997)
also verified that the method can handle multiperiodic line-profile variations.

\begin{figure}
\centering
\resizebox{0.45\textwidth}{!}{\rotatebox{0}{\includegraphics{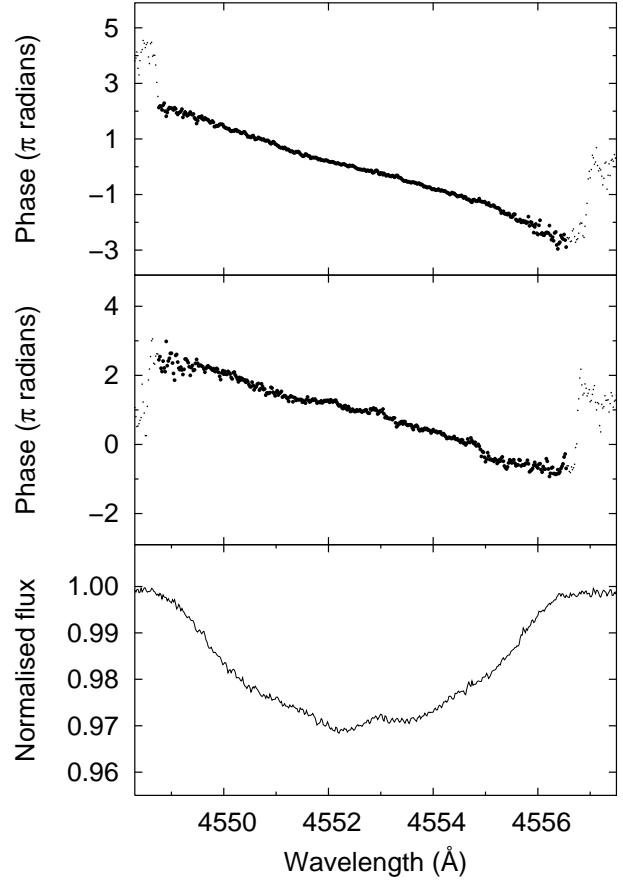}}}
\caption{Phase behaviour across the profile calculated on the basis of all CES
and CORALIE data for ($f_1$,$f_2$) = (7.415,\,4.542)\,c\,d$^{-1}$.  The upper
(middle) panel shows the phase behaviour of $f_1$ ($f_2$).  The average line
profile is given in the lower panel.}
\label{fig_cen_amp_pha_3x3}
\end{figure}

We applied this method to different combinations of alias frequencies
($f_1$,$f_2$) with $f_1 = 7.415$\,c\,d$^{-1}$ or one of its adjacent aliases,
and $f_2 = 5.546$ or one of its adjacent aliases and the results are the same
for each of them. The resulting phase behaviour for ($f_1$,$f_2$) =
(7.415,\,4.542)\,c\,d$^{-1}$ is shown in Fig.\,\ref{fig_cen_amp_pha_3x3}.  We
obtain smooth phase distributions across the line profile which allow a reliable
application of Telting \& Schrijvers' (1997) linear relations to estimate
$\ell$.
For the first frequency $f_1 = 7.415$\,c\,d$^{-1}$, we can read off a
blue-to-red phase difference $\Delta\Psi_{f_1}$ of 5$\pi$ radians, irrespective
of the value of $f_2$.  This implies that 
$\ell_1 \in [4,7]$. The phase diagram of
4.542\,c\,d$^{-1}$ leads to a phase difference $\Delta\Psi_{f_2} \in
[3.5,3.6]\,\pi$ radians, hence $\ell_2 \in [3,5]$.  These results explain why
variations in ground-based photometry are absent because $\ell$ values above
two lead to strong cancellation across the visible stellar disk in such data
(see, e.g., Dziembowski 1977).

We scanned the database of seismic $\beta\,$Cephei star models and
their oscillation frequencies computed by Ausseloos et al.\ (2004) once more,
considering
the tight limitations on the physical parameters of $\beta\,$Cen's
primary. Despite the narrow allowed range in the mass, the 
effective temperature, and the 
age of the primary, we could find numerous predicted modes with a frequency near
$f_1$ or $f_2$ for the allowed models. As we have no definitive mode degree for
the two frequencies, nor any estimate of their azimuthal orders, we cannot
refine the physical parameters of the primary from the oscillations at this
stage.

\section{Conclusions}

We have shown a systematic error to occur in the 
semi-amplitudes of the velocity curves, due to an
underestimation of radial-velocity values in SB2 spectroscopic
binaries with merged spectral line profiles.  We provide
methodological schemes to solve for this systematic error. They 
are based on spectral disentangling by means of the {\sc korel}
code (Hadrava 1997). We suggest that these schemes be used in any
future analyses of SB2s whenever their profiles can be
successfully disentangled.

In the case of $\beta\,$Cen, the systematic underestimation of the
spectroscopic orbital semi-amplitudes led to an underestimation of
the component masses of about 10\%. We refined the component
masses of this massive binary by application of our analysis
schemes to the available high-resolution spectroscopy, and by
combining the spectroscopic results with interferometric
measurements across the orbit, leading to a precision of 6\%. The
accuracy was further improved by stellar modelling taking into
account an extensive database of stellar evolution models with
wide ranging values of the mass, $Z$, and core convective
overshooting. In this way, we find the component masses of
$\beta\,$Cen to be $M_1=10.7\pm 0.1\,M_\odot$ and $M_2=10.3\pm
0.1\,M_\odot$ and its age to be $14.1\pm 0.6$ million years. These
mass estimates turn out to be fully compatible with the
mass-luminosity relation. The fact that we find $\beta\,$Cen to
have passed less than half of its main-sequence lifetime is
compatible with its high eccentricity and suggests that both
components were formed together, rather than having undergone a
tidal capture. The absence of an IR excess (Aerts et al.\ 1999),
and of H$\alpha$ emission in the CORALIE spectra, exclude the stars
still being in their pre-main-sequence phase. The {\it a
posteriori\/} conclusion that the derived fundamental parameters
of the components of the system fulfill the tight mass-luminosity
relation provides confidence in our high-precision estimates of
the masses. The determination of the distance to $\beta\,$Cen is
also affected by the previous systematic underestimation of the
radial-velocity values and has been re-determined from its dynamic
parallax to be 108$\pm$4\,pc.

Next, we performed an in-depth line-profile analysis.  All previous frequency
analyses were focused on the line-profile variations of the component producing
the deeper lines (secondary) and claimed frequencies for this
component. We have given compelling evidence that it is actually the component
producing the broader lines (primary) which is pulsating, while we did not find
any periodic variability that we could link to the secondary. As our dataset is
by far the most extensive one so far used for a spectroscopic analysis of
$\beta$\,Cen, the pulsational nature of the secondary should be
regarded as unproven. If the secondary oscillates, its amplitudes must be much
smaller than those of the primary because the variations of the combined line
profiles of both components are dominated by the variations of the primary's
line profile.

We detected two frequencies in the primary's line-profile variations by means of
the 2D Lomb-Scargle method, but we were not able to fix their value
unambiguously due to aliasing. Notwithstanding the aliasing effect, we were able
to restrict the value of the degree $\ell$ to [4,\,7] and [3,\,5]
for the first and second mode, respectively. 
The detection of only two frequencies and the lack of unique mode identification
prevented an in-depth seismic study of the star, despite the fact that its
fundamental parameters are so tightly constrained by the binarity.
The only conclusion we could draw in this respect is that standard
stellar models predict frequencies that are fully compatible with the two
detected ones.

Very few accurate masses of $\beta\,$Cephei stars are available, notable
exceptions being those with a seismic mass estimate (HD\,129929: Aerts et al.\
2003; 16\,Lac: Thoul et al.\ 2003; $\nu\,$Eri: Pamyatnykh et al.\ 2004 and
Ausseloos et al.\ 2004). The masses we derived here for $\beta\,$Cen, together
with the estimates for its effective temperature, gravity, and age, constitute a
fruitful starting point for future seismic analyses of this massive binary. 
A distinct short-event photometric variation of $\beta$\,Cen with amplitude of
about 0.04\,mag was observed by Balona (1977) on one night. No period could be
derived in these data, however. The lack of other claims of photometric
variability of the brightest among all $\beta\,$Cephei stars, despite
observational efforts (L.A.~Balona, private communication), is nicely explained
by our detection of pulsation modes with a high degree ($\ell \ge 3$). This
clearly points out that one cannot hope to find a complete frequency spectrum of
the p-modes in $\beta\,$Cephei stars from ground-based photometry alone. The
same conclusion was recently drawn for the fast rotator $\zeta$\,Oph on the
basis of high-resolution spectroscopy and MOST space photometry (Walker et al.\
2005). {
It is clear that high-precision photometric data from space are necessary
to achieve a seismic interpretation of $\beta\,$Cen.}

\begin{acknowledgements}
MA, CA, and KL are supported by the Fund for Scientific Research of Flanders
(FWO) under grant G.0332.06 and by the Research Council of the University of
Leuven under grant GOA/2003/04. We are grateful to Dr.\ P.\ Hadrava for making
his computer source codes {\sc fotel} and {\sc korel} available to us.  This
study has benefited greatly from the senior fellowship awarded to PH by the
Research Council of the University of Leuven, which allowed his three-month stay
at the Institute of Astronomy in Leuven.  PH was also supported from the
research plans J13/98: 113200004 of the 
Ministry of Education, Youth and Sports, 
AV 0Z1 003909, GA~\v{C}R~205/06/0304, and project K2043105 of the Academy of
Sciences of the Czech Republic. \end{acknowledgements}

{}

\end{document}